\documentclass[a4paper,11pt]{article}
\pdfoutput=1 

\usepackage{jcappub} 
\usepackage{array, xcolor, bibentry, longtable}
\usepackage[T1]{fontenc} 
\makeatletter
\def\@fpheader{\relax}
\makeatother

\title{\boldmath Non-minimally coupled Natural Inflation: Palatini and Metric formalism with the recent BICEP/Keck}

\author[]{Nilay Bostan}

\affiliation{Department of Physics and Astronomy, University of Iowa, \\52242, Iowa City, IA, USA }

\emailAdd{nilay-bostan@uiowa.edu}

\abstract{In this work, we show the effect of the non-minimal coupling $\xi \phi^2 R$ on the inflationary parameters by considering the single-field inflation and present the inflationary predictions of the appealing potential for the particle physics viewpoint: Natural Inflation, an axion-like inflaton which has a cosine-type 
periodic potential and the inflaton naturally emerges as a pseudo-Nambu-Goldstone boson with a spontaneously broken global symmetry. We present the inflationary predictions for this potential, $n_s$, $r$, and $\alpha=\mathrm{d}n_s/\mathrm{d}\ln k$. In addition, we assume standard thermal history after inflation, and using this, for considered potential, we show compatible regions for the $n_s$, $r$ within the recent BICEP/Keck results.}

\begin{document}
\maketitle
\flushbottom

\section{Introduction}
\label{sec:intro}
Inflation is a paradigm to clarify the isotropy, homogeneity, and flatness of the Universe, in addition, the generation of the primordial density fluctuations \cite{Guth:1980zm,Linde:1981mu,Albrecht:1982wi,Linde:1983gd}. Even though we currently have several well-established phenomenological models to explain the inflationary period, its nature is still unrevealed. Furthermore, the description of inflationary scenario is based on the slowly-rolling scalar field, $\phi$, also known as \textit{inflaton}, with a flat potential $V(\phi)$. Most of the inflationary models that have been studied in the literature so far are based on inflaton, see \cite{Martin:2013tda}. 

In addition, the following parameters are used for the inflationary predictions of the inflation models which are taken into account: \textit{spectral index}, $n_s$, \textit{tensor-to-scalar ratio}, $r$, and the \textit{running of the spectral index}, $\alpha$, these parameters can be calculated and compared to constraints of the cosmic microwave background radiation (CMBR) temperature anisotropies and polarization measurements \cite{Aghanim:2018eyx,Akrami:2018odb, BICEP:2021xfz}, which are very precise with the recent results. The recent BICEP/Keck results provide good constraints, in particular to the $r$, which tightens to $r < 0.035$ \cite{BICEP:2021xfz}. These strong constraints provide a plausible explanation for the amplitude of the primordial gravitational waves and the inflationary scale. Also, from the recent results, the $2\sigma$ constraint on the $n_s$, $n_s \in [0.957, 0.976]$. Currently, the latest constraints are not very strong on the $\alpha$ to be able to test the inflationary models, but it is expected to have some improvements with the observations from the 21 cm line, approximately the level of $\alpha=\mathcal{O} (10^{-3})$ \cite{Kohri:2013mxa,Basse:2014qqa,Munoz:2016owz}. Also, the predictions of inflationary models that are described above are constrained with the pivot scale $k_* = 0.002$ Mpc$^{-1}$.

In this work, we study the inflationary parameters for one of the very important inflation models for the particle physics viewpoint: Natural Inflation (NI). This model of inflation was first proposed by Freese, Frieman, and Olinto in 1990 for the explanation of the fine-tuning problem of inflation \cite{Freese:1990rb}. The inflation potential has to be nearly flat and NI models provide a plausible solution for this. Natural Inflation is utilizing the \textit{axionic field} that has flat inflaton potential due to the shift symmetry. With this, the inflaton in this model is a Nambu-Goldstone boson with a very flat potential that inflation and slow-roll expansion require. The Natural Inflation model has an axion-like inflaton which has a cosine-type periodic potential. In addition, several modified NI models have been suggested, such as \cite{German:2021jer}. In this work, we will focus on the original cosine type of the Natural Inflation potential.

We take into account the inflationary parameters for this potential, $n_s$, $r$, and $\alpha=\mathrm{d}n_s/\mathrm{d}\ln k$ with the assumption of the standard thermal history after inflation and for this potential, we show compatible regions for the $n_s$, $r$ within the recent BICEP/Keck results. In addition, we consider the non-minimal coupling $\xi \phi^2 R$ between the Ricci scalar and inflaton, which is required for the renormalizable scalar field theory \cite{Callan:1970ze, Freedman:1974ze, Buchbinder:1992rb} in curved space-time. It is important to note that depending on the non-minimal coupling function $\xi$, inflationary parameters can change significantly, and accordingly, whether inflation will occur or not may change \cite{Abbott:1981rg,Spokoiny:1984bd,Lucchin:1985ip,Futamase:1987ua,Fakir:1990eg,Salopek:1988qh,Amendola:1990nn,Faraoni:1996rf,Faraoni:2004pi}.
In literature, a large number of studies take into account the non-minimally coupled inflation either in Metric or Palatini formulation \cite{Bezrukov:2010jz, Bostan:2018evz, Bezrukov:2007ep,Tenkanen:2017jih,Bauer:2008zj,Rasanen:2017ivk,Jinno:2019und,Rubio:2019ypq,Enckell:2018kkc, Bostan:2019wsd, Tenkanen:2019jiq,Jarv:2020qqm}. In general relativity (GR), the metric and its first derivatives are the independent variables for the Metric formulation \cite{Padmanabhan:2004fq,Paranjape:2006ca}. On the other hand, the metric and the connection are independent in the Palatini formulation \cite{Attilio, Einstein, Ferraris}. For the Einstein-Hilbert Lagrangian, equations of motion are identical for two formulations, it means that they actually describe the same physical theories. However, if there is a non-minimal coupling between matter and gravity, these two formulations cannot be equivalent, therefore they describe two different theories of gravity (\cite{Bauer:2008zj,York:1972sj,Tenkanen:2017jih,Rasanen:2017ivk,Racioppi:2017spw,Tamanini:2010uq}). Some examples of the differences between Metric and Palatini formulations are as follows: for large $\xi$ values, the $\xi$-attractor models appear in the Starobinsky model in the Metric formulation, on the other hand, this behavior does not appear in Palatini formalism \cite{Kallosh:2013tua, Jarv:2017azx}. Also, for large $\xi$ values, $r$ can take very tiny values in the Palatini formulation compared to the Metric formalism \cite{Racioppi:2017spw,Barrie:2016rnv,Kannike:2015kda,Artymowski:2016dlz}. Furthermore, in the literature, another approach for the inflation that can be considered is the affine approach, \textit{Affine Inflation}, see \cite{Azri:2017uor, Azri:2018qux}. Affine gravity \cite{Kijowski:2004bj} is based upon the connection with no notion of metric. This framework requires the scalar fields which have non-vanishing potentials, therefore it is considered that Affine gravity is important while studying inflation.

The paper is organized as follows: we first describe the non-minimally coupled inflation for the Metric and Palatini formulations (section \ref{general}), we then discuss Natural Inflation potential in detail and show our results for this potential with non-minimal coupling with the assumption of the standard thermal history after inflation (section \ref{NI}). Finally, in section \ref{conc}, we discuss and conclude our results in the paper.

\section{Non-minimally coupled inflation: Metric and Palatini formulations}\label{general}

We first describe the form of Jordan frame action 

\begin{eqnarray}\label{nonminimal_action}
S_J = \int \mathrm{d}^4x \sqrt{-g}\Big(\frac{1}{2}F(\phi) g^{\mu\nu}R_{\mu\nu}(\Gamma) - \frac{1}{2} g^{\mu\nu}\partial_{\mu}\phi\partial_{\nu}\phi -V_J(\phi)\Big),
\end{eqnarray}
here, $g_{\mu \nu}$ is a determinant of the space-time metric and $J$ indicates that the action is defined in the Jordan frame. $\phi$ is the inflaton and $V_J(\phi)$ is the potential defined in the Jordan frame. Also, $F(\phi)$ is a non-minimal coupling function, $R_{\mu\nu}$ is the Ricci tensor and it is given as follows
\begin{equation}\label{Riccitensor}
R_{\mu\nu}=\partial_{\sigma}\Gamma_{\mu \nu}^{\sigma}-\partial_{\mu}\Gamma_{\sigma \nu}^{\sigma}+\Gamma_{\mu \nu}^{\rho}\Gamma_{\sigma \rho}^{\sigma}-\Gamma_{\sigma \nu}^{\rho}\Gamma^{\sigma}_{\mu \rho}.
\end{equation}
The connection is described with the metric tensor function in the Metric formulation. It is the Levi-Civita connection ${\bar{\varGamma}={\bar{\varGamma}}(g^{\mu\nu})}$
\begin{equation} \label{vargammametric}
\bar{\varGamma}_{\mu\nu}^{\lambda}=\frac{1}{2}g^{\lambda \rho} (\partial_{\mu}g_{\nu \rho}+\partial_{\nu}g_{\rho \mu}-\partial_{\rho}g_{\mu\nu}).
\end{equation}
On the other hand, the connection $\varGamma$ and $g_{\mu \nu}$ are taken as independent variables in the Palatini formulation and with the assumption that the torsion-free connection, i.e. $\varGamma_{\mu\nu}^{\lambda}=\varGamma_{\nu\mu }^{\lambda}$. After solving the equations of motion, one can obtain the form \cite{Bauer:2008zj}
\begin{eqnarray}\label{vargammapalatini}
\Gamma^{\lambda}_{\mu\nu} = \overline{\Gamma}^{\lambda}_{\mu\nu}
+ \delta^{\lambda}_{\mu} \partial_{\nu} \omega(\phi) +
\delta^{\lambda}_{\nu} \partial_{\mu} \omega(\phi)
- g_{\mu \nu} \partial^{\lambda} \omega(\phi),
\end{eqnarray}
and here
\begin{eqnarray}
\label{omega}
\omega\left(\phi\right)=\ln\sqrt{\frac{F(\phi)}{M^2_{P}}},
\end{eqnarray}
where $M_{P}=(8 \pi G)^{-1/2}$. We will utilize the units, with the reduced Planck scale $M_P=1/\sqrt{8\pi
G}\approx2.4\times10^{18}\text{ GeV}$ will be set equal to unity.

After inflation, $F(\phi)\to1$ or $\phi\to0$. In this paper, we will compute the inflationary predictions by considering the non-minimally coupled Natural Inflation potential.  We use $F(\phi)=1+\xi \phi^n$ \cite{Reyimuaji:2020goi} for the non-minimal coupling function, here $n$ takes even numbers.

\subsection{A brief description of inflationary parameters}
Calculating the inflationary predictions in the Einstein frame is useful. We can switch from the Jordan frame to Einstein frame ($E$) by using Weyl rescaling, $g_{E, \mu \nu}=g_{\mu \nu}/F(\phi)$. It is important to note that in the Einstein frame, all matter couplings become $\phi$ dependent ($g_E$ metric) and thus, during the reheat phase at the end of inflation, the inflaton can decay into matter with a rate determined by $F(\phi)$ since $g_{\mu\nu} = F(\phi) g_{E, \mu \nu}$. 

Einstein frame action is 
\begin{eqnarray}\label{einsteinframe}
S_E = \int \mathrm{d}^4x \sqrt{-g_{E}}\Big(\frac{1}{2}g_E^{\mu\nu}R_{E, \mu \nu}(\Gamma)-\frac{1}{2Z(\phi)}\, g_E^{\mu\nu} \partial_{\mu}\phi\partial_{\nu}\phi - V_E(\phi) \Big),
\end{eqnarray}
where
\begin{eqnarray} \label{Zphi}
Z^{-1}(\phi)=\frac{3}{2}\frac{F'(\phi)^2}{F(\phi)^2}+\frac{1}{F(\phi)} \rightarrow Metric, \ \ \ Z^{-1}(\phi)=\frac{1}{F(\phi)} \rightarrow Palatini, 
\end{eqnarray}
\begin{equation} \label{Zphi2}
V_E(\phi)=\frac{V_J(\phi)}{F(\phi)^2},
\end{equation}
$F'\equiv\mathrm{d}F/\mathrm{d}\phi$. If the field redefinition is made by using,
\begin{equation}\label{redefine}
\mathrm{d}\chi=\frac{\mathrm{d}\phi}{\sqrt{Z(\phi)}},
\end{equation}
Einstein frame action can be found for the minimally coupled scalar field $\chi$ with the canonical kinetic term. After using field redefinition, Einstein frame action in terms of $\chi$ is
\begin{eqnarray}\label{einsteinframe2}
S_E = \int \mathrm{d}^4x \sqrt{-g_{E}}\Big(\frac{1}{2}g_E^{\mu\nu}R_{E}(\Gamma)
-\frac{1}{2}\, g_E^{\mu\nu} \partial_{\mu}\chi\partial_{\nu}\chi - V_E(\chi) \Big).
\end{eqnarray}
Provided that the Einstein frame potential is written in terms of the canonical scalar field $\chi$, by utilizing slow-roll parameters, inflationary predictions can be found \cite{Lyth:2009zz}. The slow-roll parameters in terms of $\chi$ are as follows
\begin{equation}\label{slowroll1}
\epsilon =\frac{M^2_{P}}{2}\left( \frac{V_{\chi} }{V}\right) ^{2}\,, \quad
\eta = M^2_{P}\frac{V_{\chi\chi} }{V}  \,, \quad
\zeta ^{2} = M^4_{P} \frac{V_{\chi} V_{\chi \chi\chi} }{V^{2}}\,,
\end{equation}
here the subscripts $\chi$'s indicate derivatives. In the slow-roll approximation, inflationary parameters can be found in the form
\begin{eqnarray}\label{nsralpha1}
n_s = 1 - 6 \epsilon + 2 \eta \,,\quad
r = 16 \epsilon, \nonumber\\
\alpha=\frac{\mathrm{d}n_s}{\mathrm{d}\ln k} = 16 \epsilon \eta - 24 \epsilon^2 - 2 \zeta^2\,.
\end{eqnarray}
In the slow-roll approximation, the number of e-folds can be found using the form
\begin{equation} \label{efold1}
N_*=\frac{1}{M^2_{P}}\int^{\chi_*}_{\chi_e}\frac{V\rm{d}\chi}{V_{\chi}}\,. \end{equation}
Here, the subscript ``$_*$'' denotes the quantities at the scale corresponding to $k_*$ exited the horizon. Furthermore, $\chi_e$ is the value of inflaton at the end of inflation, we can calculate this by using $\epsilon(\chi_e) = 1$. The number of e-folds is around 60, however, the exact value should depend upon the evolution of the Universe, which we will describe in the next part of this paper. 

The curvature perturbation amplitude in terms of $\chi$ can be found
\begin{equation} \label{perturb1}
\Delta_\mathcal{R}^2=\frac{1}{12\pi^2 M^6_{P}}\frac{V^3}{V_{\chi}^2},
\end{equation}
which should be well-matched $\Delta_\mathcal{R}^2\approx 2.1\times10^{-9}$ from the Planck results \cite{Aghanim:2018eyx} for the pivot scale $k_* = 0.002$ Mpc$^{-1}$.

In addition to this, for our numerical calculations, we need to have the slow-roll parameters in terms of the original field, $\phi$, so we need to reformulate the slow-roll parameters that we defined above. The reason why we need this redefinition is that calculating the inflationary potential in terms of $\chi$ for general $\xi$ and $v$ values is not simple. Utilizing eq. \eqref{redefine}, eq. \eqref{slowroll1} can be obtained in terms of $\phi$ \cite{Linde:2011nh}
\begin{eqnarray}\label{slowroll2}  
\epsilon=Z\epsilon_{\phi}\,,\quad
\eta=Z\eta_{\phi}+{\rm sgn}(V')Z'\sqrt{\frac{\epsilon_{\phi}}{2}}, \nonumber \\
\zeta^2=Z\left(Z\zeta^2_{\phi}+3{\rm sgn}(V')Z'\eta_{\phi}\sqrt{\frac{\epsilon_{\phi}}{2}}+Z''\epsilon_{\phi}\right),
\end{eqnarray}
where we define 
\begin{equation}
\epsilon_{\phi} =\frac{1}{2}\left( \frac{V^{\prime} }{V}\right) ^{2}\,, \quad
\eta_{\phi} = \frac{V^{\prime \prime} }{V}  \,, \quad
\zeta ^{2} _{\phi}= \frac{V^{\prime} V^{\prime \prime\prime} }{V^{2}}\,.
\end{equation}
Also, eqs. \eqref{efold1} and \eqref{perturb1} can be found in terms of $\phi$
\begin{eqnarray}\label{perturb2}
N_*&=&\rm{sgn}(V')\int^{\phi_*}_{\phi_e}\frac{\mathrm{d}\phi}{Z(\phi)\sqrt{2\epsilon_{\phi}}}\,,\\
\label{deltaR} \Delta_\mathcal{R}&=&\frac{1}{2\sqrt{3}\pi}\frac{V^{3/2}}{\sqrt{Z}|V^{\prime}|}\,.
\end{eqnarray}

Moreover, we will assume the standard thermal history after inflation, and considering this, we will calculate the inflationary parameters for the Natural Inflation potential. With this assumption, $N_*$ is in the form \cite{Liddle:2003as}
\begin{eqnarray} \label{efolds}
N_*\approx64.7+\frac12\ln\frac{\rho_*}{M^4_{P}}-\frac{1}{3(1+\omega_r)}\ln\frac{\rho_e}{M^4_{P}} 
+\Big(\frac{1}{3(1+\omega_r)}-\frac14\Big)\ln\frac{\rho_r}{M^4_{P}}\,,
\end{eqnarray}
here, $\rho_{e}=(3/2)V(\phi_{e})$, the
energy density at the end of inflation. $\rho_r$ indicates the energy density at the end of reheating, and $\rho_{*} \approx V(\phi_*)$ is the energy density when the scales corresponding to $k_*$ exited the horizon and $\rho_*$ can be expressed by using eq. \eqref{deltaR} as follows
\begin{equation}
    \rho_{*} = \frac{3 \pi^2\Delta^2_\mathcal{R} r }{2},
\end{equation}
here, $r$ is the tensor-to-scalar ratio, $\Delta^2_\mathcal{R}$, the amplitude of curvature perturbation, which we already defined above. In addition to this, $\omega_r$ is the equation of state parameter during reheating. We will take $\omega_r$ as constant. 

We can assign three reasonable ranges for $N_*$ which depends upon the reheating temperature ($T_r$): 

\begin{itemize}
\item \textit{High-$N$ case:} $\omega_r=1/3$, which corresponds to the assumption of instant reheating.
\item \textit{Medium-$N$ case:} $\omega_r=0$ and $T_r=10^{9}$ GeV and $\rho_r$ can be calculated by utilizing the Standard Model value of the number of relativistic degrees of freedom, $g_*=106.75$. 
\item \textit{Low-$N$ case:} $\omega_{r}=0$, which is the same as the medium-$N$ case but the reheat temperature is $T_r=100$ GeV. 
\end{itemize}
 \begin{figure}[tbp]
\centering
\includegraphics[width=.47\textwidth]{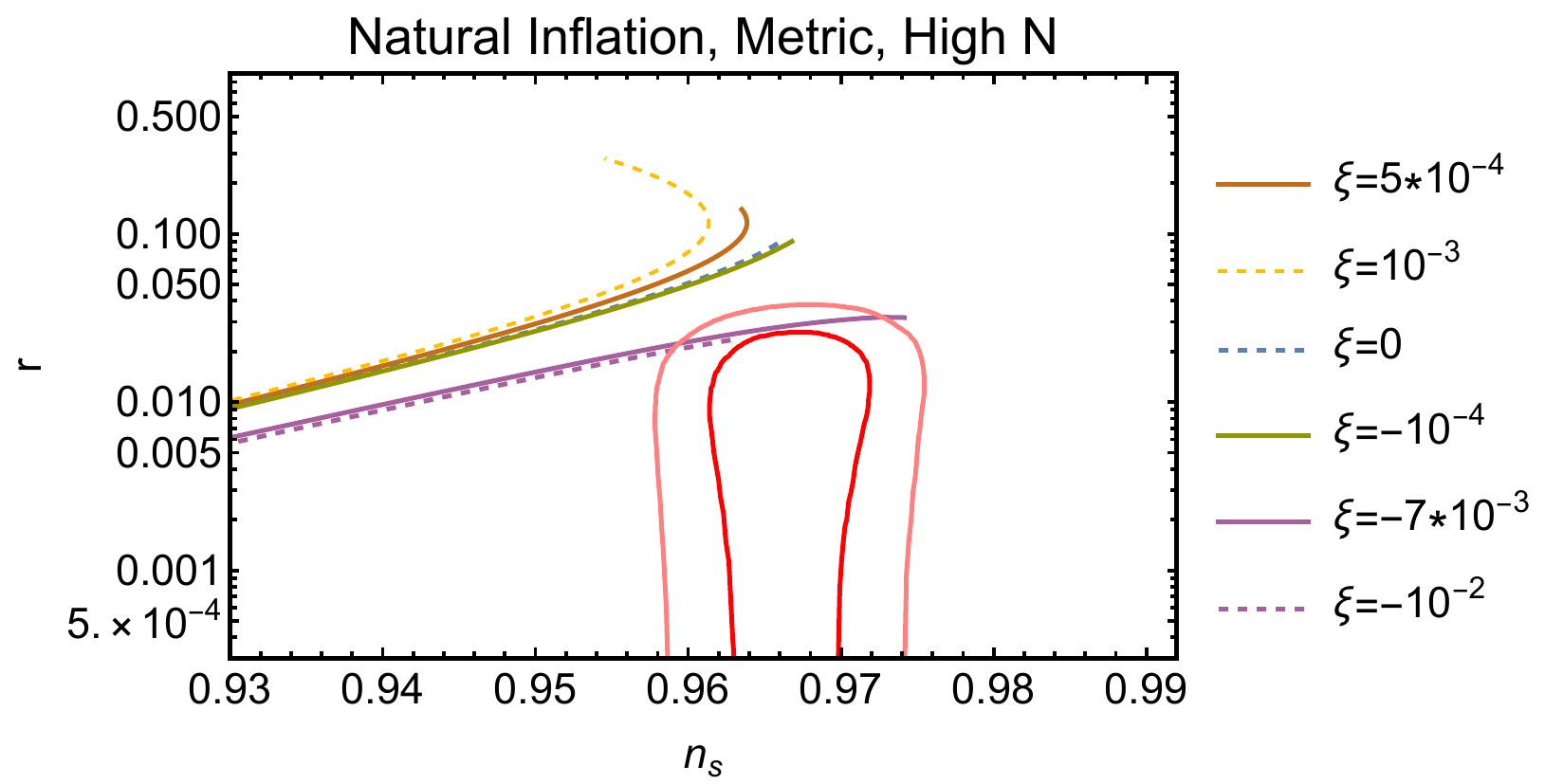}
\includegraphics[width=.47\textwidth]{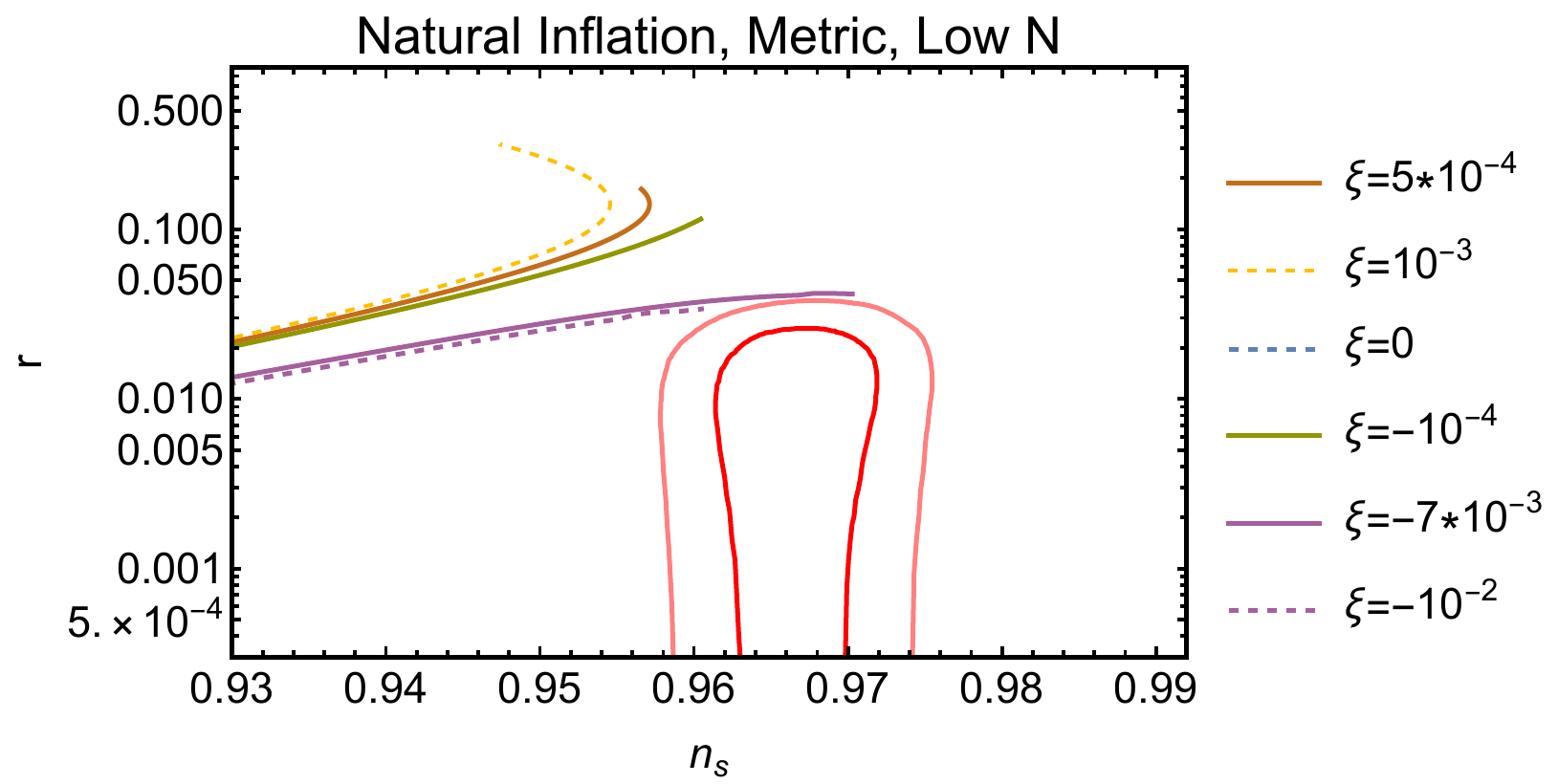}
\caption{\label{n2met} The inflationary predictions of the Natural inflation potential in Metric formulation for $n=2$. Figures show $n_s-r$ for chosen $\xi$ values for high-$N$ (left) and low-$N$ (right) cases. The pink (red) contours show the 95\% (68\%) CL from the recent BICEP/Keck \cite{BICEP:2021xfz}.}
\end{figure}
\section{Natural Inflation}\label{NI}
Natural inflation was first proposed to be able to solve the problem called \textit{fine-tuning} of inflation \cite{Martin:2013tda}. Also, Natural Inflation provides a reasonable solution to the flatness of inflaton potential that is required. It is described by the axion-like potential \cite{Freese:1990rb,Adams:1992bn,Reyimuaji:2020goi}, and it is such an appealing model in particle physics because the inflaton field, $\phi$, naturally emerges being a pseudo-Nambu-Goldstone boson from the spontaneously broken global symmetry, shift symmetry \cite{Freese:1990rb}. Moreover, $\phi$ is the axion-like inflaton that has a cosine-type periodic potential. The form of the Natural Inflation potential in the Jordan frame is defined as
\begin{equation}\label{nipot}
V_J(\phi)=V_0\left[1+ \cos \left(\frac{\phi}{f}\right)\right].
\end{equation}
\begin{figure}[tbp]
\centering
\includegraphics[width=.45\textwidth]{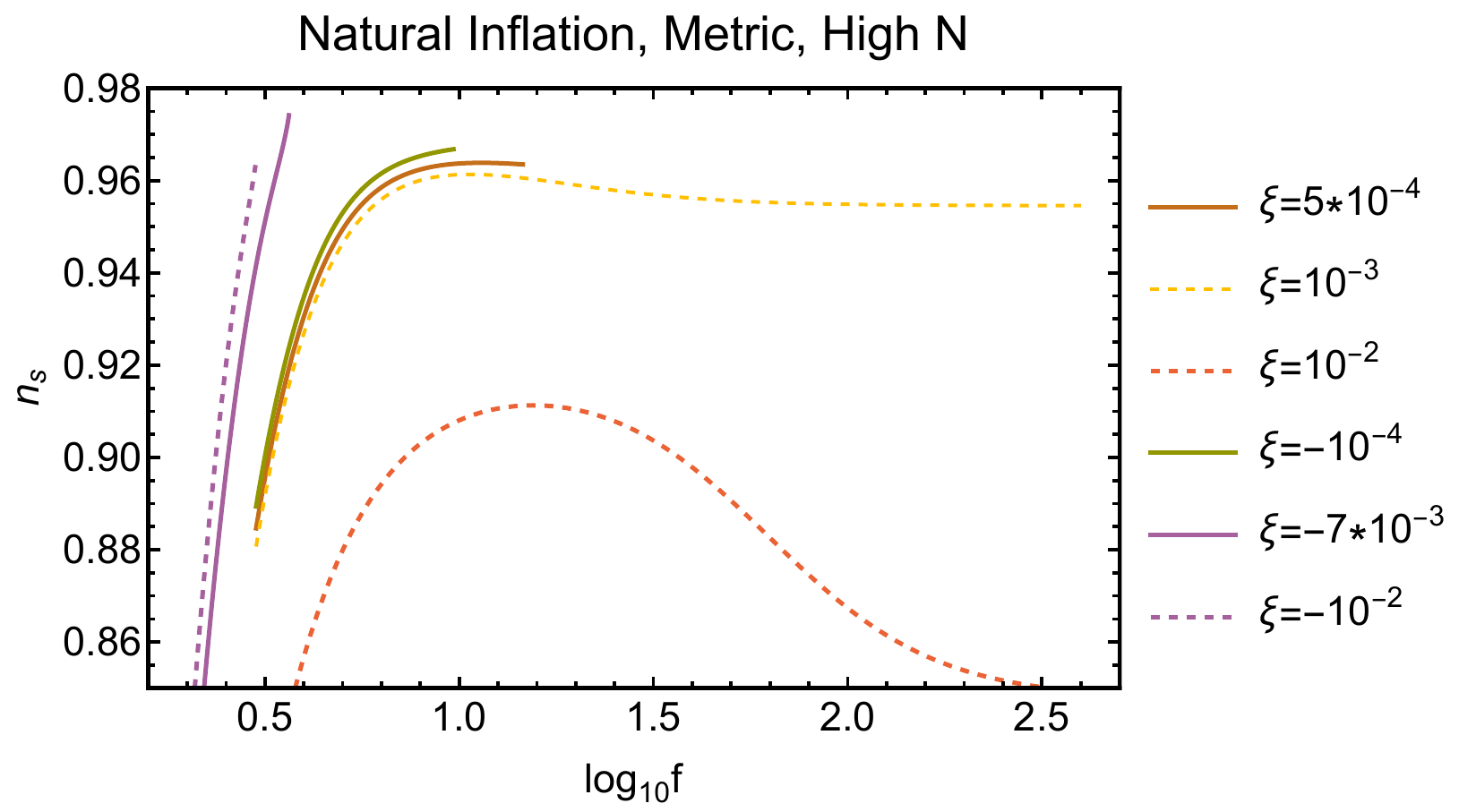}
\includegraphics[width=.45\textwidth]{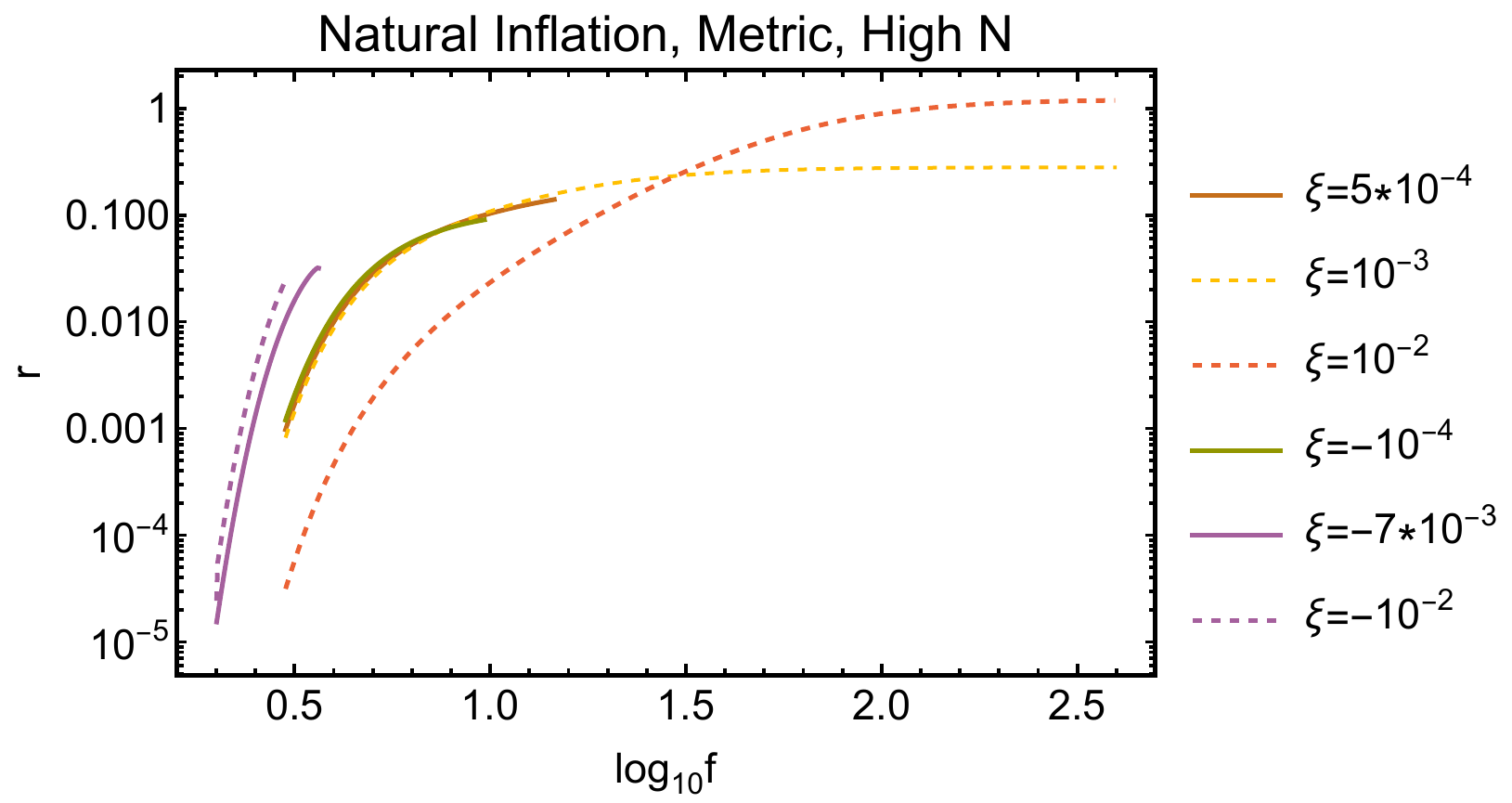}
\caption{\label{n2met2} For the Natural inflation potential, figures show the predictions of $n_s$, $r$ for chosen $\xi$ values as a function of $f$, considering the high-$N$ case, Metric formulation and $n=2$.}
\end{figure}
Here, $f$ is the symmetry-breaking scale. In this section, we will study the inflationary parameters of non-minimally coupled Natural Inflation for both Metric and Palatini formulations, we use the assumption of instant reheating, which we defined as the High-$N$ case, and for the low-$N$ case that corresponds to $\omega_{r}=0$ and $T_r=100$ GeV. We compare our results with the recent BICEP/Keck data. 
\begin{figure}[tbp]
\centering
\includegraphics[width=.45\textwidth]{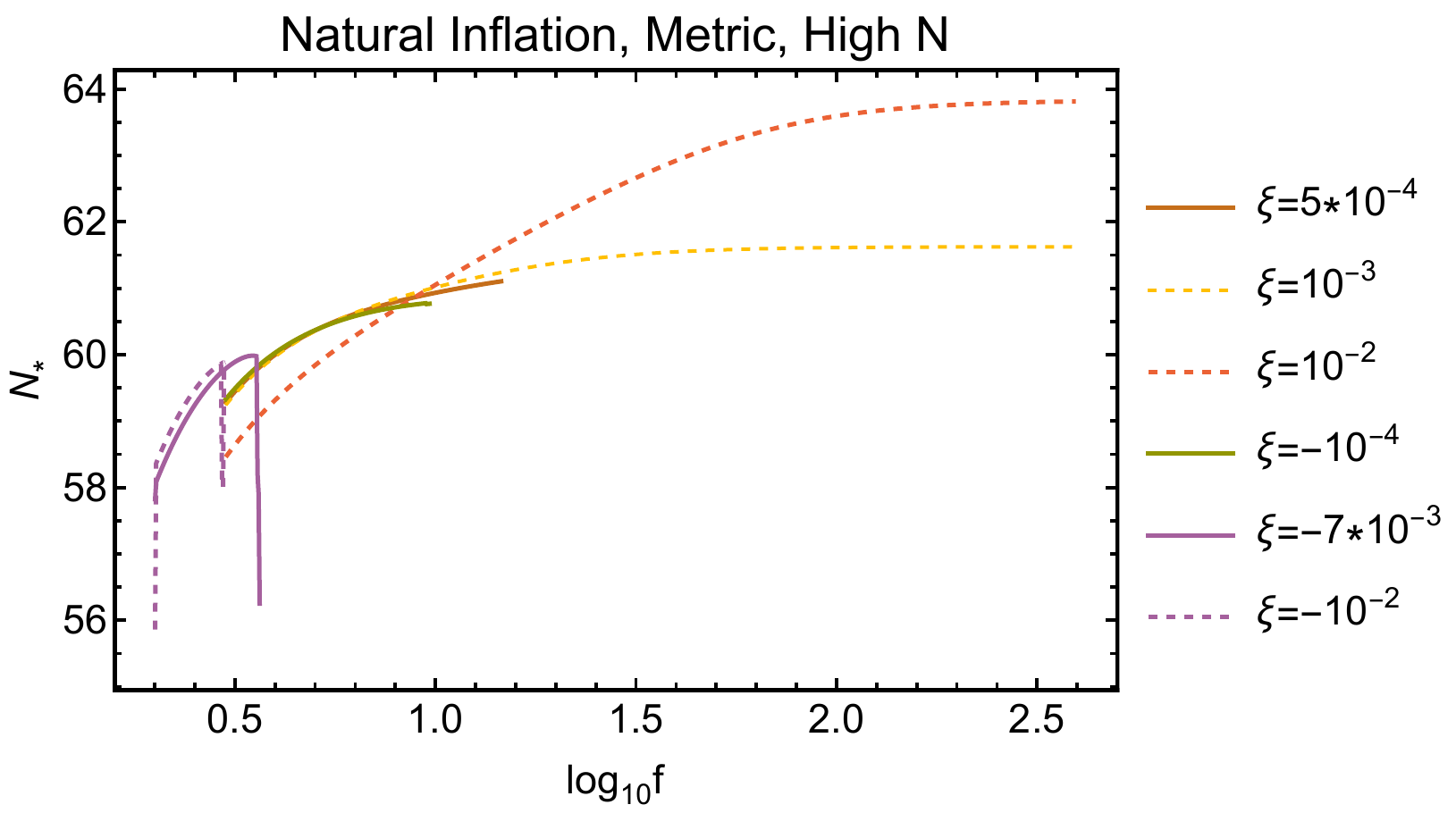}
\includegraphics[width=.45
\textwidth]{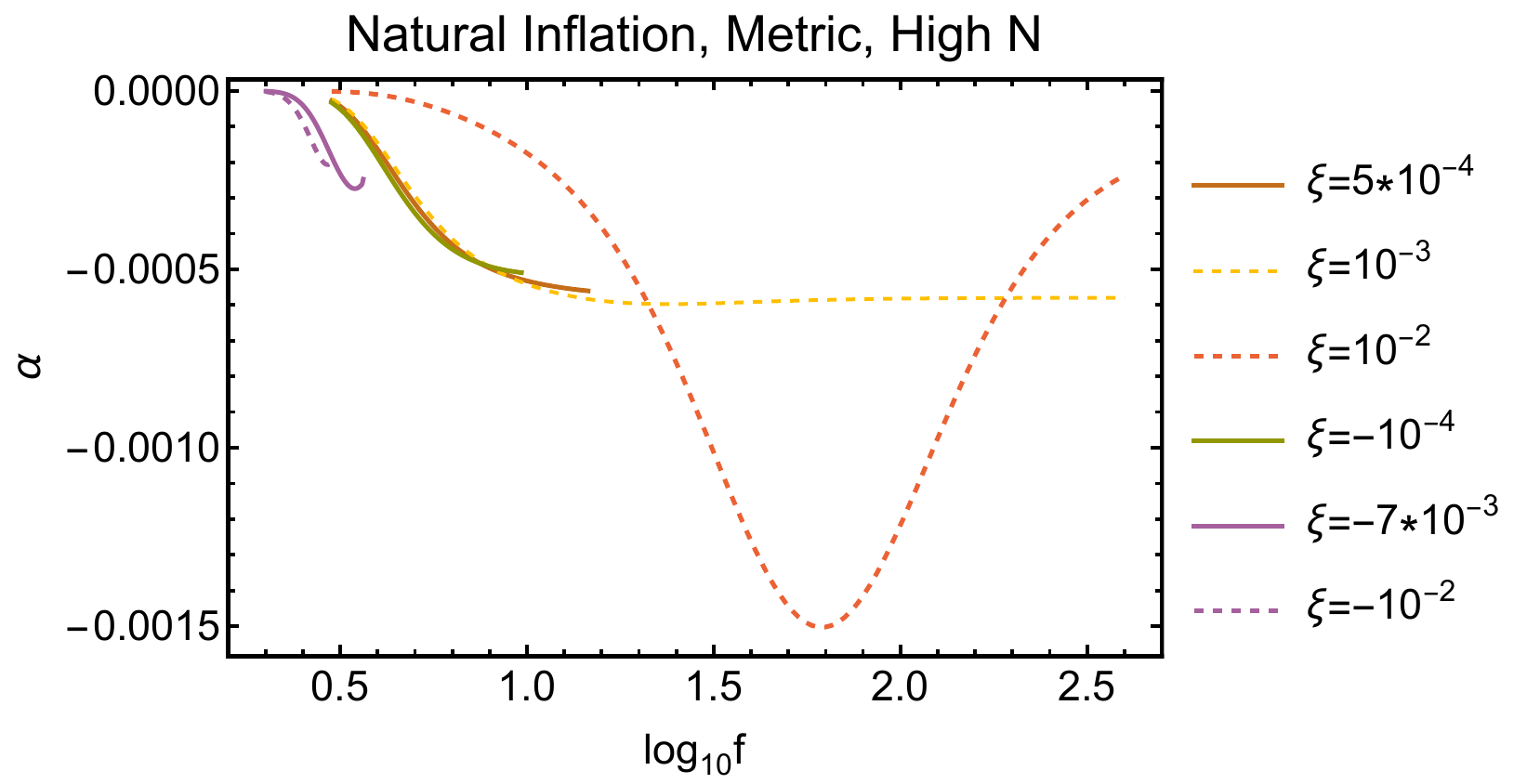}
\caption{\label{n2met3} For the Natural inflation potential, figures show the predictions of $N_*$, $\alpha$ for chosen $\xi$ values as a function of $f$, considering the high-$N$ case, Metric formulation and $n=2$.}
\end{figure}
We can write the non-minimally coupled Natural Inflation potential in Einstein frame as follows
\begin{equation}\label{nipot2}
V_E(\phi)=\frac{V_0\left[1+ \cos \left(\frac{\phi}{f}\right)\right]}{\Big(1+\xi \phi^n\Big)^2}.
\end{equation}
Also, this Einstein frame potential can be written in terms of the canonical scalar field $\chi$ inserting $\phi(\chi)$ to this potential form. In the large-field limit, $\xi \phi^2 \gg 1$, one can find $\phi(\chi)$ for $n=2$ (lowest) case for Metric and Palatini formulations by using eqs. \eqref{Zphi} and \eqref{redefine} as follows
\begin{eqnarray} \label{phipot}
\phi\simeq \frac{1}{\sqrt{\xi}}\exp{\left(\frac{\chi}{\sqrt{6}}\right)} \rightarrow in \ Metric,  \ \
\phi \simeq \frac{1}{\sqrt{\xi}}\sinh\left({\chi \sqrt{\xi}}\right) \rightarrow in \ Palatini,
\end{eqnarray} 
on condition that $\xi \gg 1$. Also, for the weak coupling limit ($\xi \phi^2\ll1$), $\phi\approx \chi$ and $V_J(\phi)\approx V_E(\chi)$. In addition to this, due to the form of the potential that we consider, there are no analytical solutions for $n>2$ cases. Moreover, we can find the $n_s$ and $r$ analytically for $n=2$ and look at different limits to simplify the expressions. As an example, let us write $r$ for the Einstein frame Natural Inflation potential defined in eq. \eqref{nipot2} and $n=2$ case
\begin{equation*}
   r\simeq \frac{8\left(4 \xi \phi (1+\cos{(\frac{\phi}{f})})+\frac{(1+\xi \phi^2)\sin{(\frac{\phi}{f})}}{f}\right)^2}{\left(1+\xi \phi^2(1+6\xi)\right)(1+\cos(\frac{\phi}{f}))^2} \rightarrow Metric,
\end{equation*}
\begin{equation}
   r\simeq \frac{8\left(4 \xi \phi (1+\cos{(\frac{\phi}{f})})+\frac{(1+\xi \phi^2)\sin{(\frac{\phi}{f})}}{f}\right)^2}{\left(1+\xi \phi^2\right)(1+\cos(\frac{\phi}{f}))^2} \rightarrow Palatini.
\end{equation}

One can find that for the limit of $6\xi\ll1$ (weak coupling), the analytical form of $r$ should be equivalent for two formulations. Thus, $r$ values should be the same in the weak coupling limit for the Metric and Palatini formulations.
\begin{figure}[tbp]
\centering
\includegraphics[width=.50\textwidth]{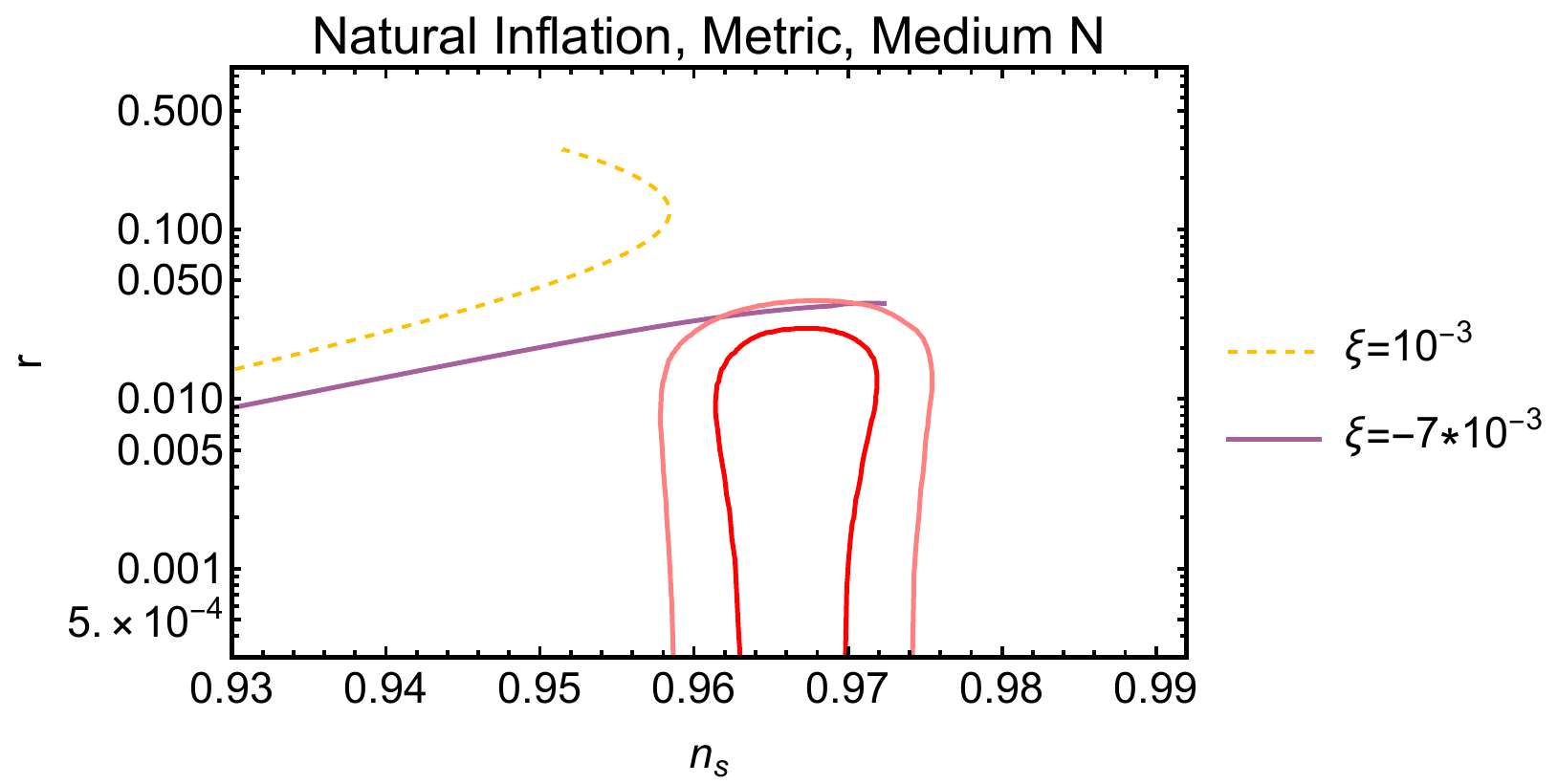}
\caption{\label{medN} The inflationary predictions of the Natural inflation potential in Metric formulation for $n=2$. Figure shows $n_s-r$ for chosen $\xi$ values for medium-$N$ case. The pink (red) contours show the 95\% (68\%) CL from the recent BICEP/Keck \cite{BICEP:2021xfz}.}
\end{figure}

In literature, Natural Inflation has been examined in several studies for both minimal ($\xi=0$) and non-minimal couplings ($\xi \neq 0$) so far, such as \cite{Reyimuaji:2020goi, Zhou:2022ovp, Stein:2021uge,German:2021jer,Simeon:2020lkd}. In \cite{Reyimuaji:2020goi}, non-minimally coupled Natural Inflation was examined for the Metric and Palatini formulations with 60 e-folds and in the range of 50-60 e-folds before the end of inflation, chosen with $F(\phi)=1+\xi \phi^n$. They showed their results by fixing $f$($\xi$) but changing $\xi(f)$. They concluded that $n=2$ (lowest case) gives the best consistency with the cosmological observations from the Planck 2018 results \cite{Aghanim:2018eyx,Akrami:2018odb}. Also, they found that $f$ should be larger than $1.9 (1.95)$ in the Metric(Palatini) formulations in order to be in agreement with Planck 2018. Moreover, for the Metric approach, $n=2$ and $f<1.9$, $n_s$ and $r$ stay outside the observations from Planck 2018. Also, $2 \leq f \leq 7.7$, the predictions can be in both $95\%$ and $68\%$ CL regions. They observed that $n_s$ decreases when $\xi$ increases. For $n=4$, smaller $\xi$ gives a better consistency with the observations. In addition to this, they showed that because the inflaton couples to gravity weakly, two formulations give very similar results. They also showed that Natural Inflation with $\xi=0$ is in good agreement with Planck 2018 results only for $f\gtrsim 5$.

Also, another important study on Natural Inflation is \cite{Zhou:2022ovp}. They discussed three different single-field modified Natural Inflation potential models with reheating constraints and showed their results for different $\omega_{r}$ and $f$ values and compared them with the recent BICEP/Keck results. Furthermore, in \cite{German:2021jer}, they discussed a modification of the Natural Inflation potential with $f<1$ (in Planck units) and they showed their results of the inflationary predictions in the minimal coupling case. 

Unlike existing studies in the literature, we will show our results for the non-minimally coupled Natural Inflation for both Metric and Palatini approaches assuming instant reheating, where $\omega_{r}=1/3$ and one of selected $N_*$ cases, that is low-$N$, where $\omega_{r}=0$ and $T_r=100$ GeV. We will present the inflationary parameters for this potential and compare our results with the cosmological observations from the recent BICEP/Keck \cite{BICEP:2021xfz}. \\

\

Let us discuss our results in the paper:

\begin{figure}[tbp]
\centering
\includegraphics[width=.45\textwidth]{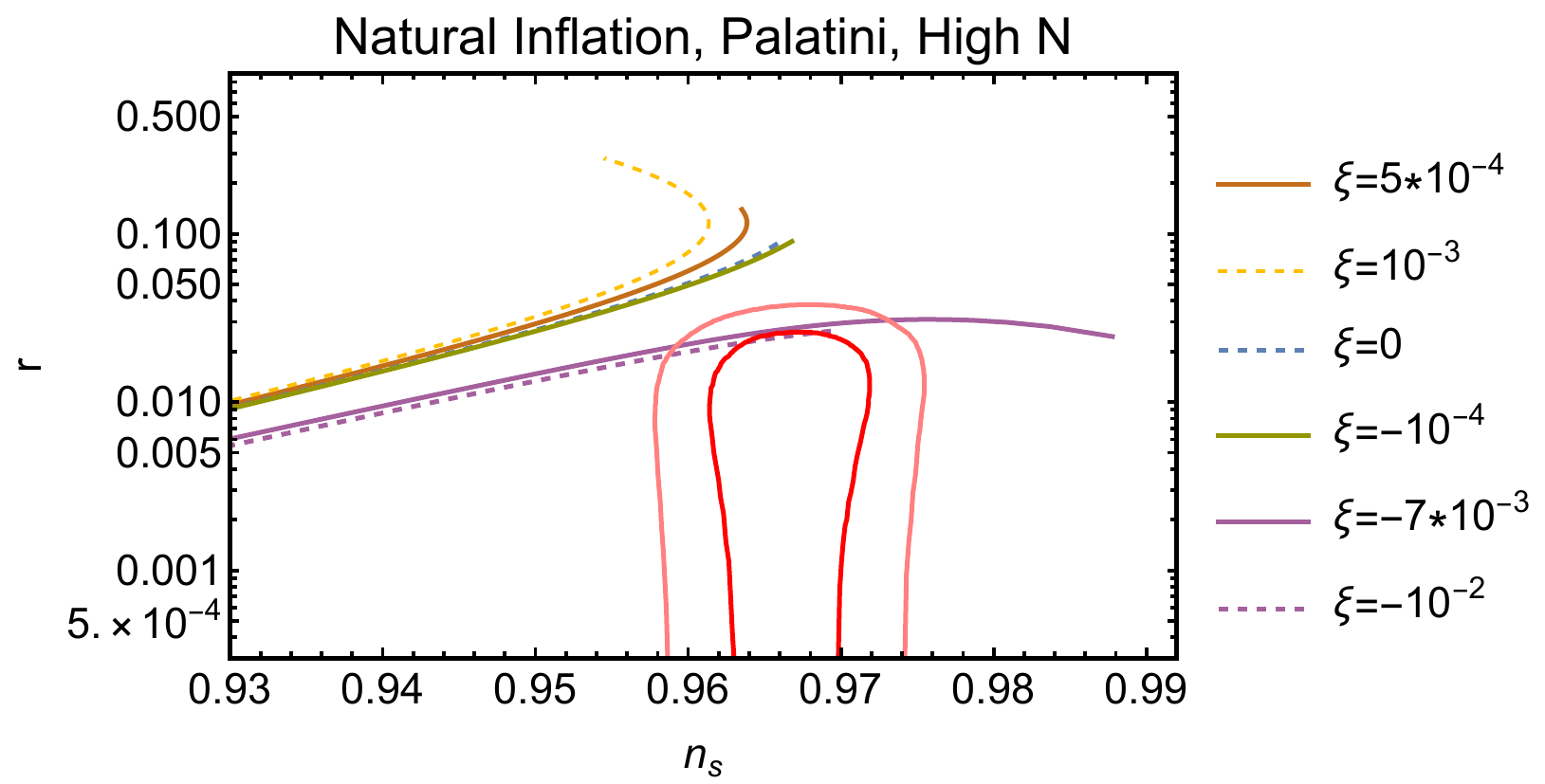}
\includegraphics[width=.45\textwidth]{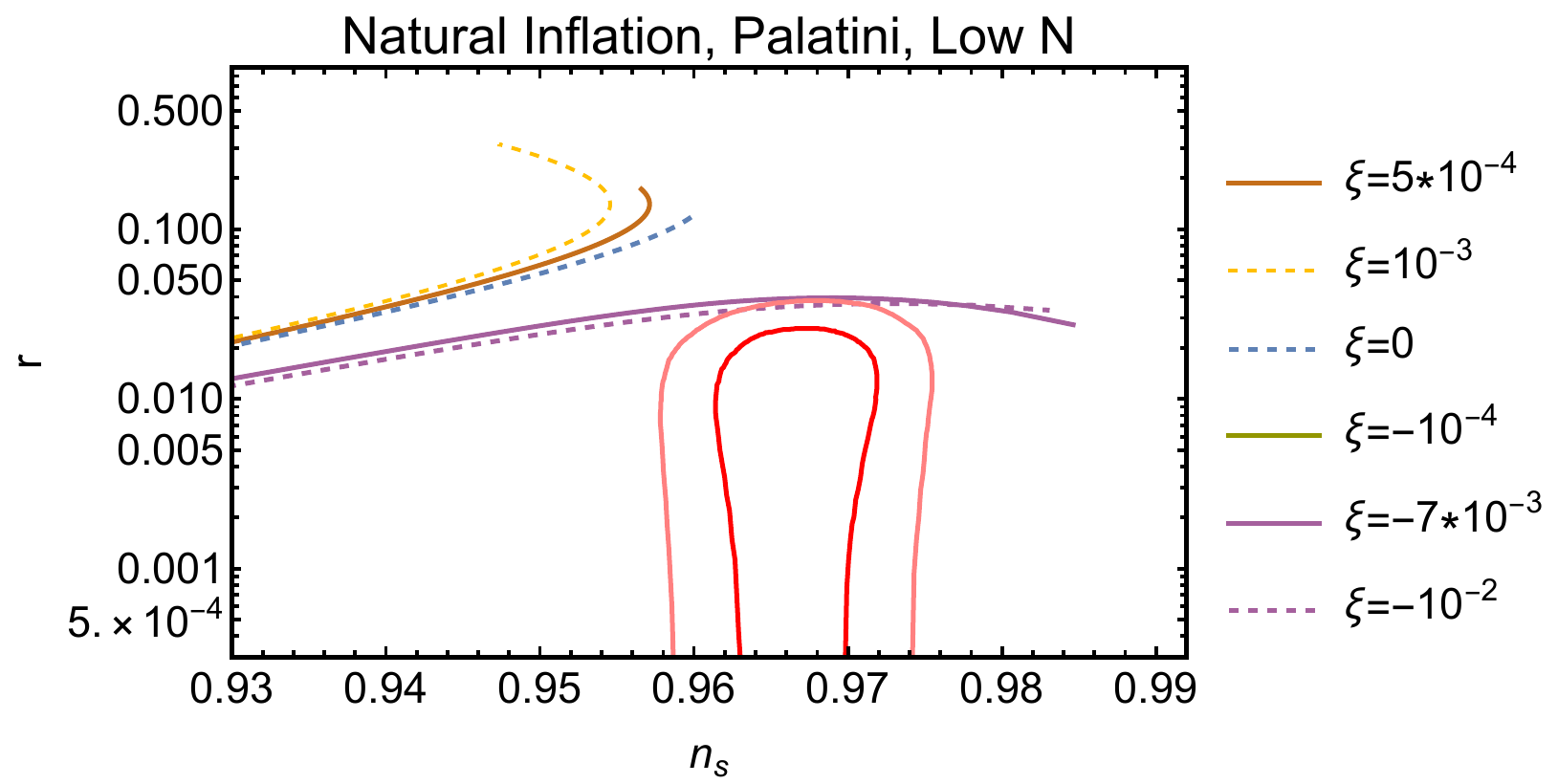}\
\caption{\label{n2pal} The inflationary predictions of the Natural inflation potential in Palatini formulation for $n=2$. Top figures show $n_s-r$ for chosen $\xi$ values for high-$N$ (left) and low-$N$ (right) cases. The pink (red) contours show the 95\% (68\%) CL from the recent BICEP/Keck \cite{BICEP:2021xfz}.}
\end{figure}
\begin{itemize}
 \item  \textit{For $n=2$ case:}

$\rightarrow$ We find that for the selected $\xi$ values, at any $f$ value, the predictions cannot be inside the observational regions for the low-$N$ case and Metric formalism. But increasing $T_{r}$, the predictions can stay in the $95\%$ CL region for $\xi=-7\times 10^{-3}$ and $f\sim 3.2$ in the Medium-$N$ case, see fig. \ref{medN}. On the other hand, for the low-$N$ case and Palatini formalism, the predictions are in the $95\%$ CL for $\xi=-10^{-2}$ and $f\sim3.2$ values. In addition, for both Metric and Palatini approaches in high-$N$ case, $n_s-r$ can enter to $95\%$ CL for the values of $\xi=-7\times 10^{-3}$ and $\xi=-10^{-2}$ at $f\sim 3.2$. Also, $N_* \sim 58$ for $\xi=-10^{-2}$ and $N_* \sim 56$ for $\xi=-7\times 10^{-3}$ when the inflationary predictions are inside the $95\%$ CL region. It is important to emphasize that the agreement with the latest constraints is only possible with $\xi<0$ values for our model. This might be considered a problem because $F(\phi)>0$ requires any arbitrary value of $\phi$. However, we only take into account negative $\xi$ values that give $F>0$ during and after the inflationary era for our model.
\begin{figure}[tbp]
\centering
\includegraphics[width=.45\textwidth]{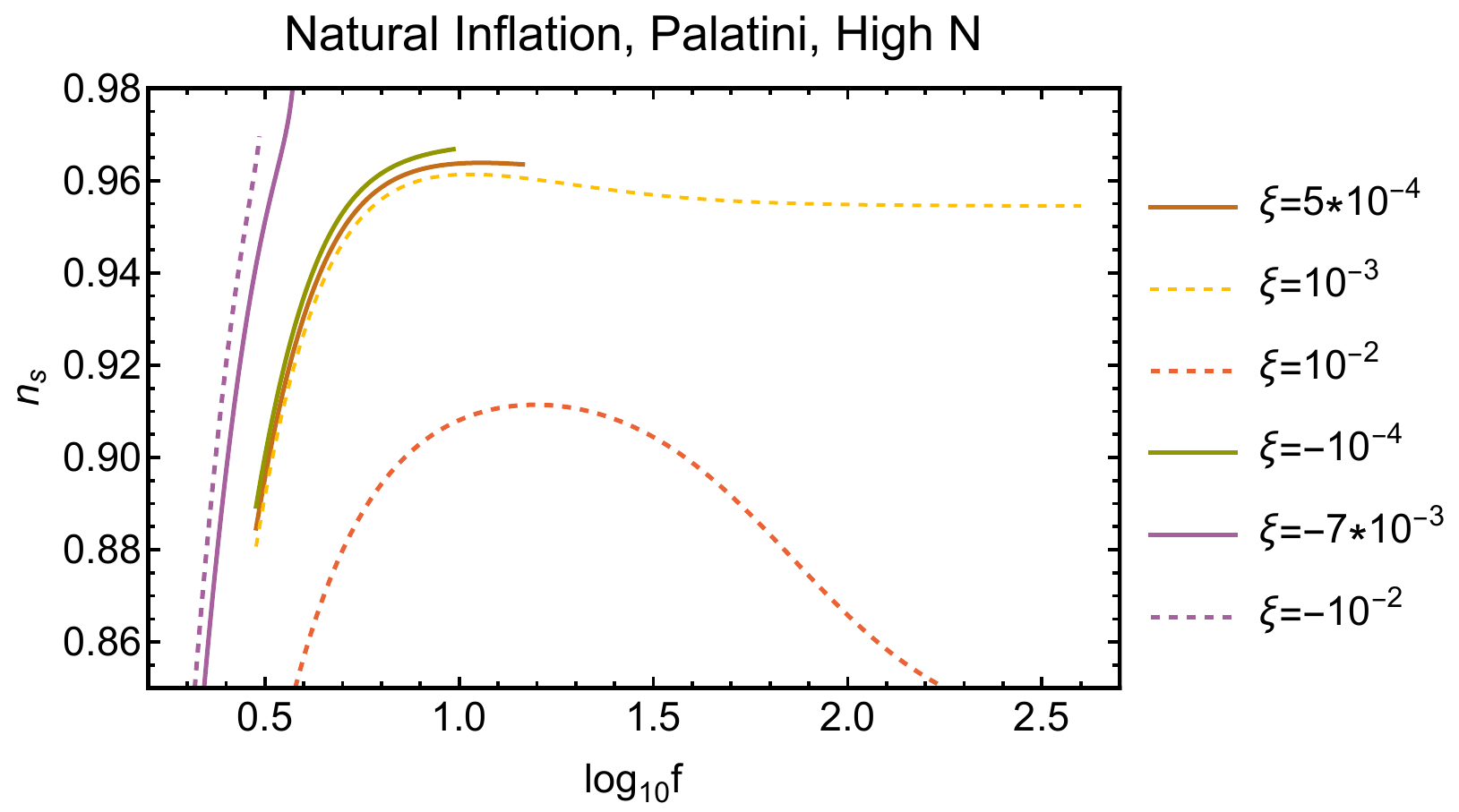}
\includegraphics[width=.45\textwidth]{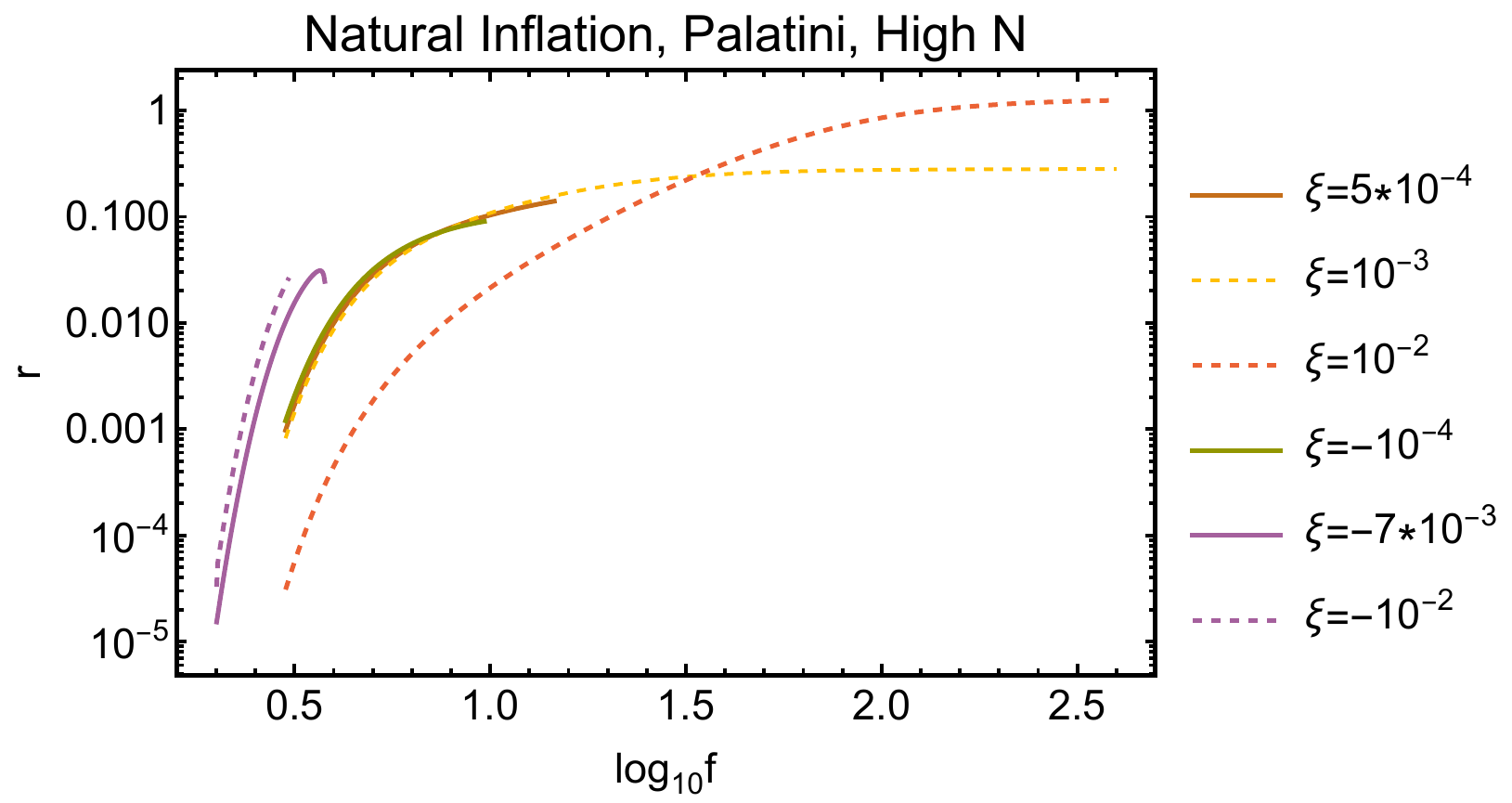}
\caption{\label{n2pal2} For the Natural inflation potential, figures show the predictions of $n_s$, $r$ as a function of $f$ for chosen $\xi$ values, considering the high-$N$ case, Palatini formulation and $n=2$.}
\end{figure}

$\rightarrow$ It is good to mention that for the Palatini formulation and high-$N$ case, for $\xi=-10^{-2}$, the predictions can be inside $68\%$ CL at $f\sim3$. Moreover, for selected $\xi$ values and both two formulations, $r$ takes very large values for any values of $f>3$, thus the predictions are ruled out with the recent observations. Also, especially for $\xi=10^{-2}$, $n_s$ values are very far away from being able to enter the CL regions for both formulations at any $f$ values. The predictions of the minimal coupling case ($\xi=0$) are ruled out for both formulations. In addition, we observed that there is a sudden decrease for $N_*$ values at $f\sim3$ for $\xi=-7\times 10^{-3}$ and $\xi=-10^{-2}$ values, this behavior is the same for two formulations with small differences. Lastly, we can conclude that the predictions for $\alpha$ are very small to be observed in the near future.
\begin{figure}[tbp]
\centering
\includegraphics[width=.45\textwidth]{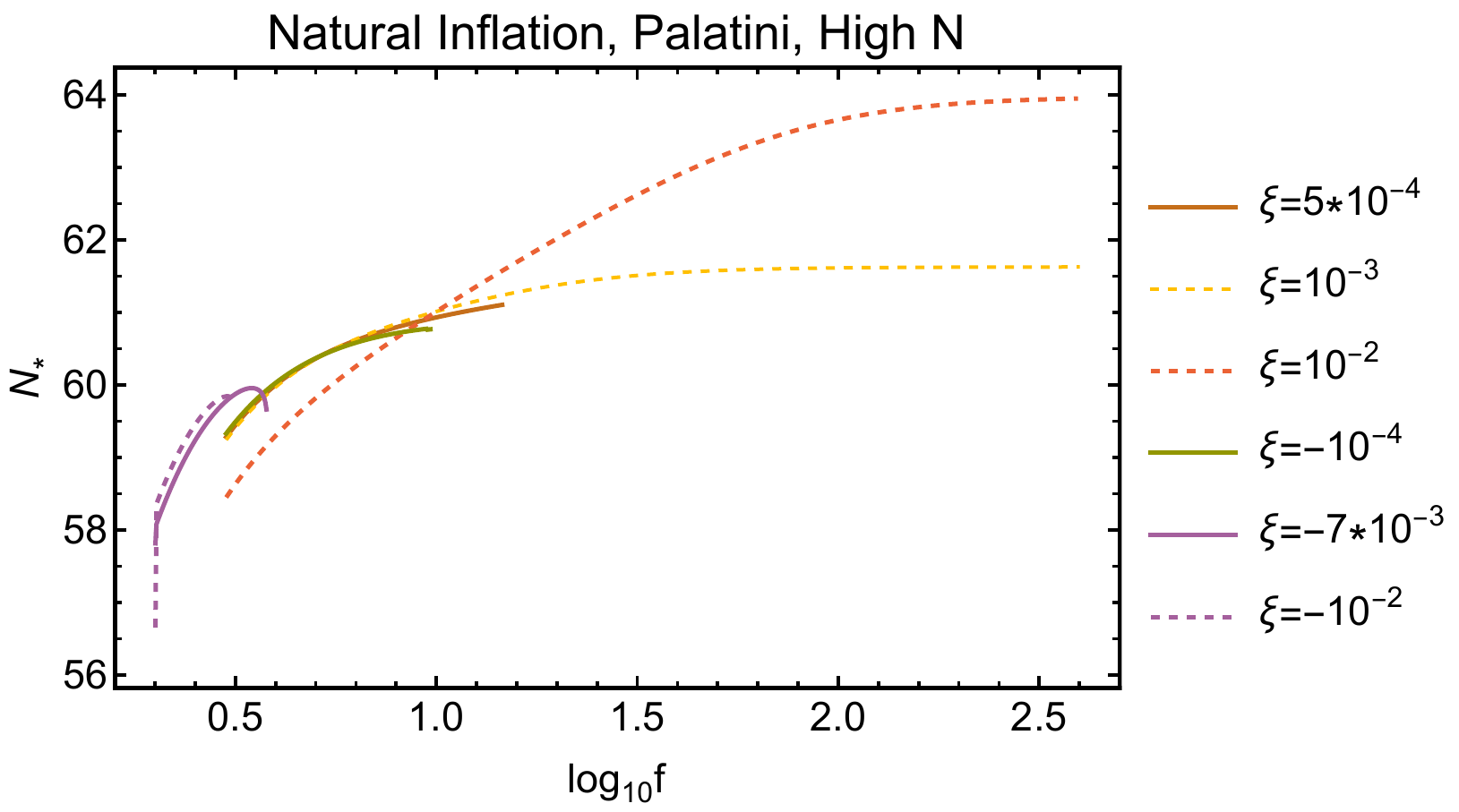}
\includegraphics[width=.45
\textwidth]{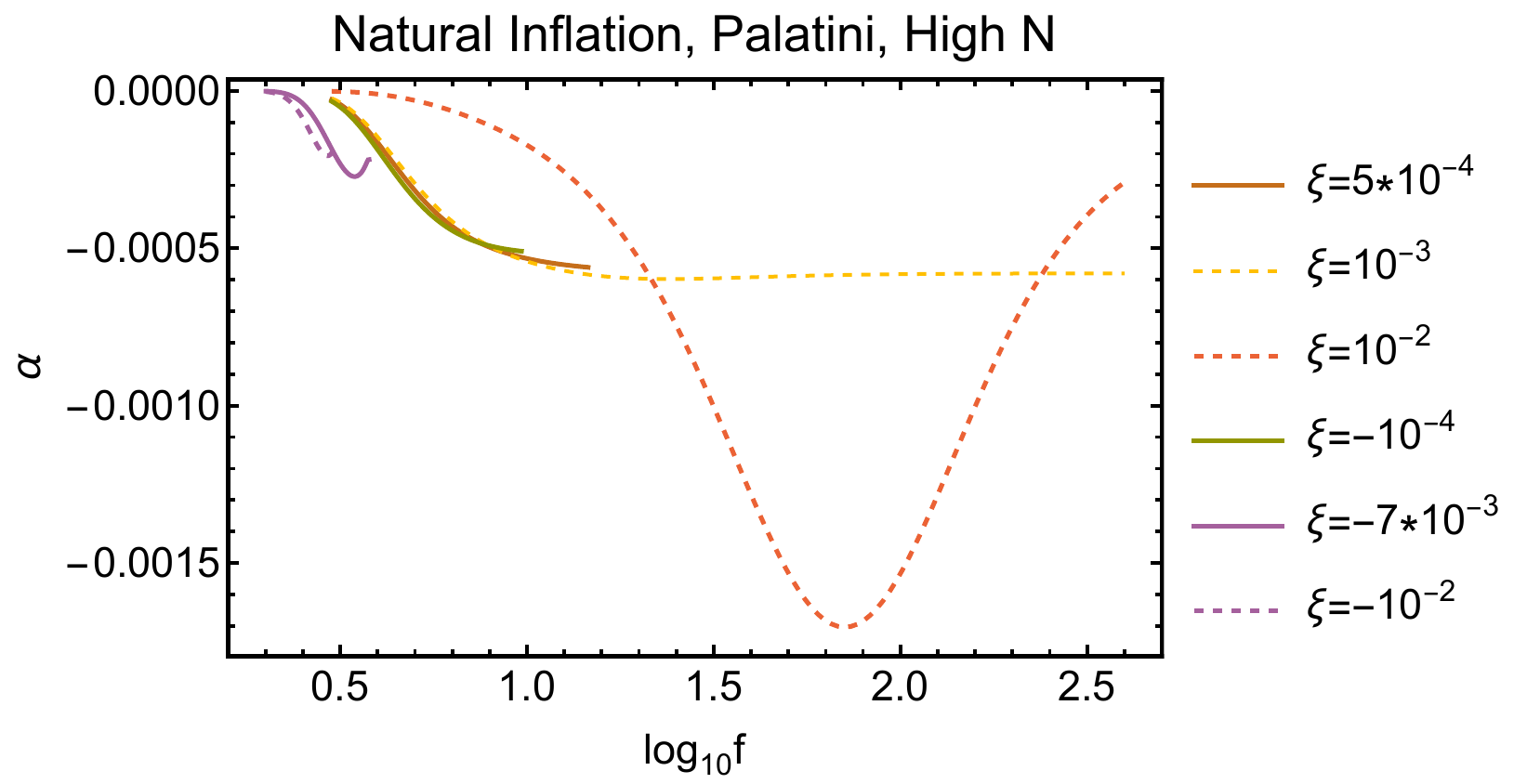}
\caption{\label{n2pal3} For the Natural inflation potential, figures show the predictions of $N_*$, $\alpha$ as a function of $f$ for chosen $\xi$ values, considering the high-$N$ case, Palatini formulation and $n=2$.}
\end{figure}

 \item  \textit{For $n=4$ case:}

The results are worse than the results of $n=2$ and the inflationary predictions are ruled out for the $\xi$ values that we selected for both Metric and Palatini formulations. For $\xi=10^{-7}$, the predictions are exactly the same for two formulations, however, for $\xi=10^{-5}$ and $\xi=10^{-3}$, the predictions are not exactly the same for two formulations, there are small differences between them for the $n_s-r$ values. Also, for $\xi=-10^{-7}$, $n_s$ and $r$ values increase as $f$ increases, at $f\sim10$, $n_s\sim 0.965$, $r\sim 0.1$ and $N_*\sim 60.5$, so the predictions are ruled out.
\begin{figure}[tbp]
\centering
\includegraphics[width=.6
\textwidth]{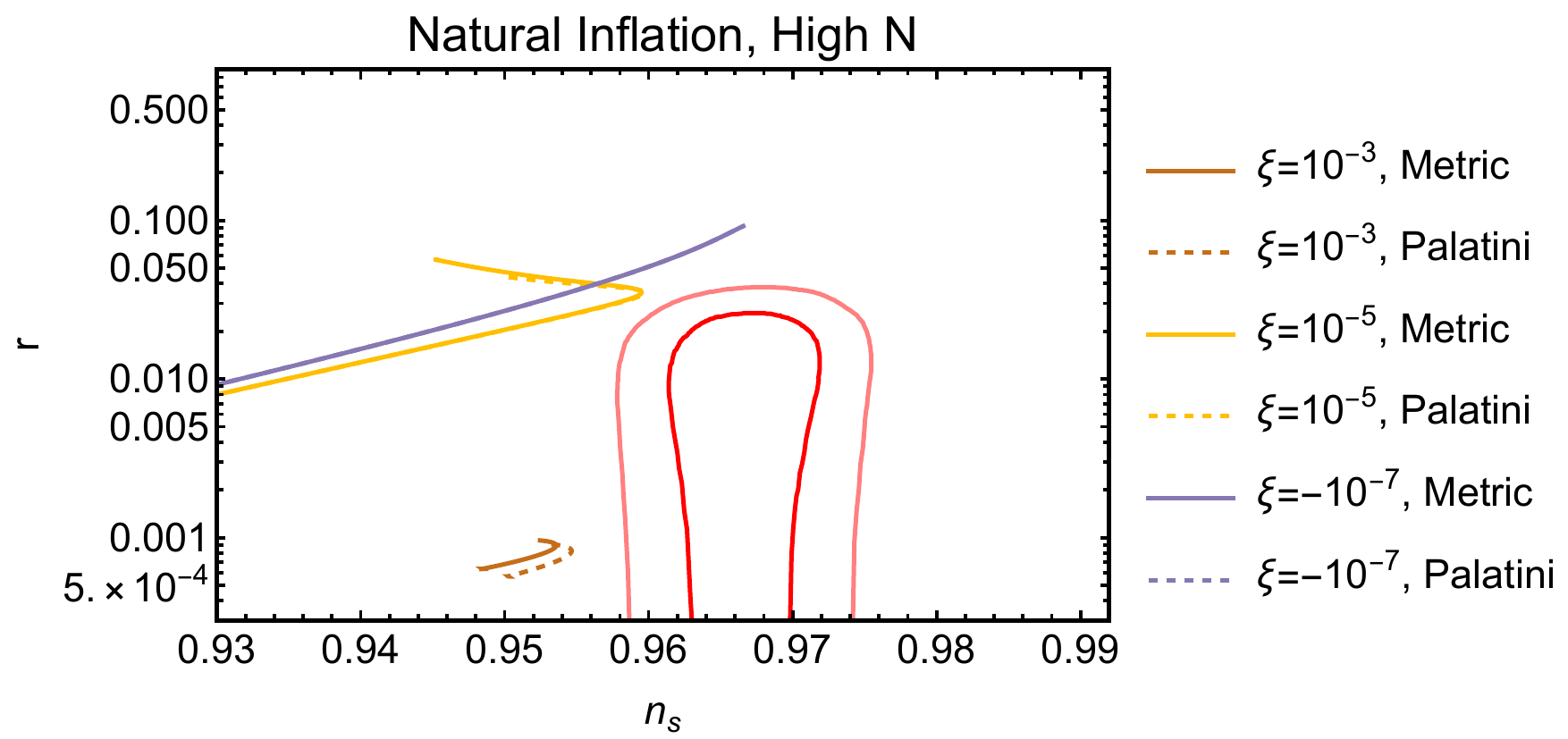}
\caption{\label{n4} The inflationary predictions of the Natural inflation potential for both formulations and $n=4$. Figures show $n_s-r$ for chosen $\xi$ values for high-$N$ case. The pink (red) contours show the 95\% (68\%) CL from the recent BICEP/Keck \cite{BICEP:2021xfz}.}
\end{figure}

\item For $\xi>0$ and $n=2$ case, when $\xi$ and $f$ values increase, $n_s$ decreases and $r$ increases, thus the predictions are naturally ruled out. So, we eliminate the inflationary predictions for the $\xi$ values larger than $10^{-2}$ in this work. This behavior is the same for both formulations. 

\item For $\xi>0$ and $n=4$ case, $n_s-r$ do not change so much while increasing $f$, also the predictions remain constant for the values of $f\gg5$. $n_s\sim 0.95$ for selected $\xi > 0$ values. Also, for $\xi=10^{-3}$ $\rightarrow$ $r \sim 5\times 10^{-4}$ and for $\xi=10^{-5}$ $\rightarrow$ $r \sim 0.04$. So, the predictions are ruled out. This behavior is also the same for both formulations.  
\begin{figure}[tbp]
\centering
\includegraphics[width=.45\textwidth]{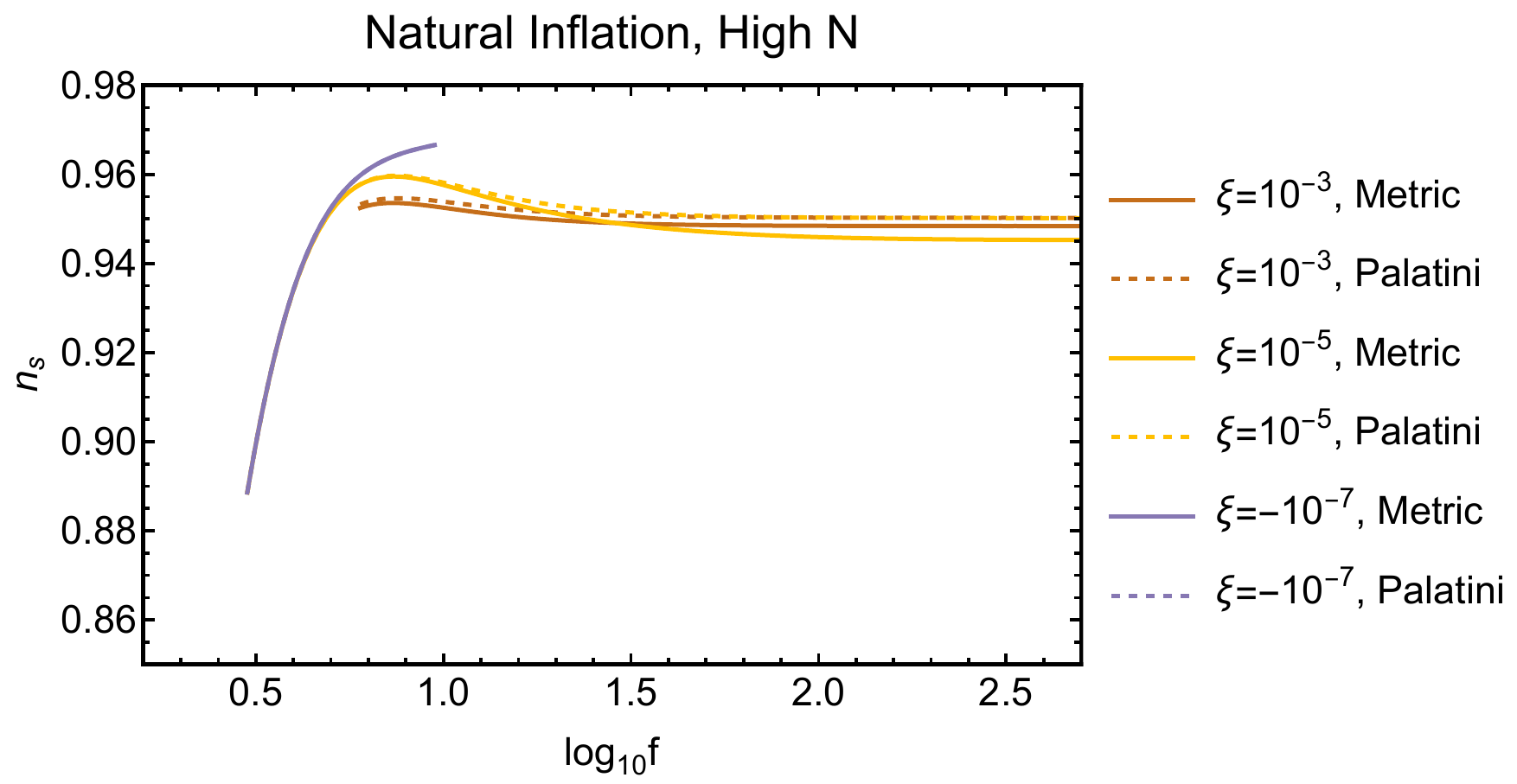}
\includegraphics[width=.45\textwidth]{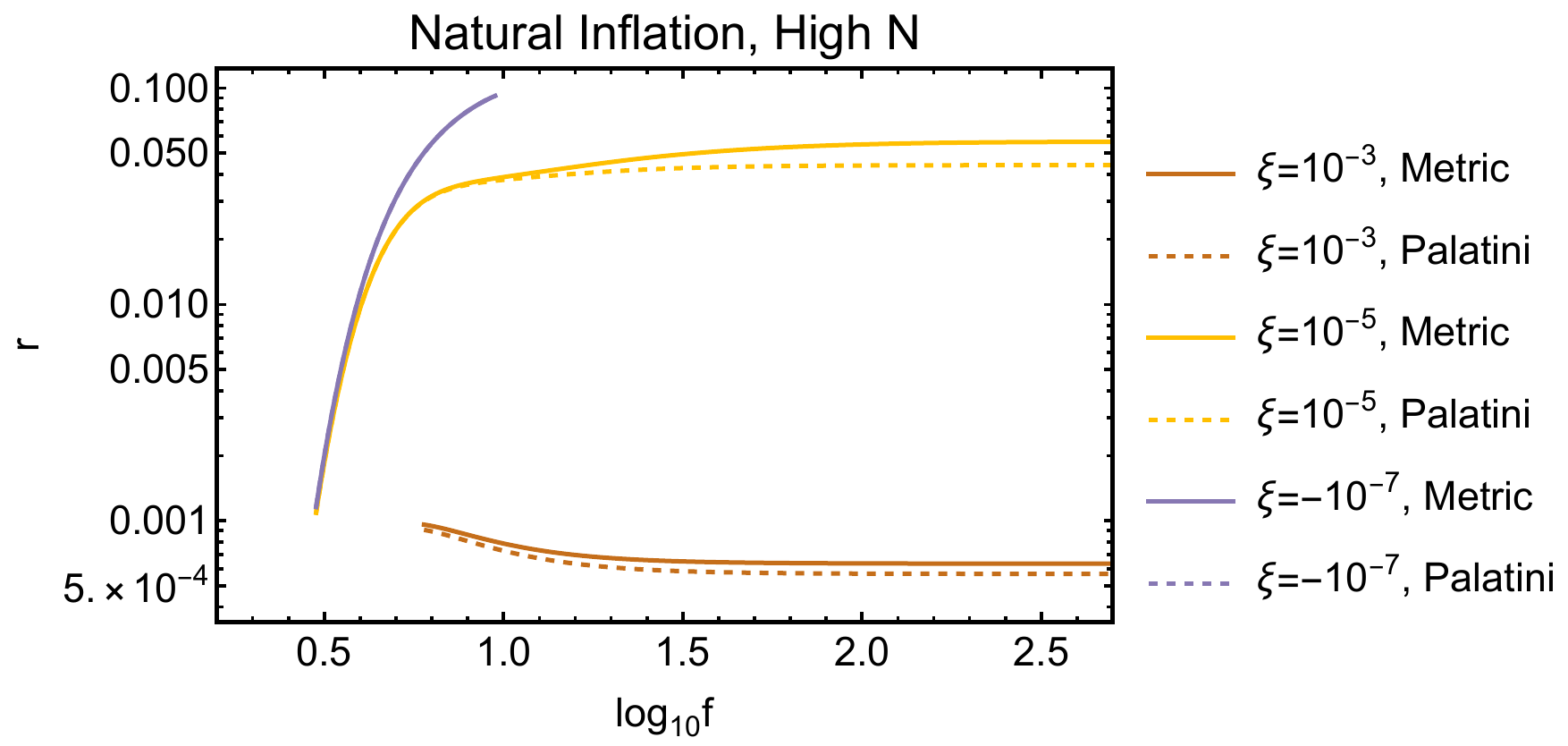}
\includegraphics[width=.45\textwidth]{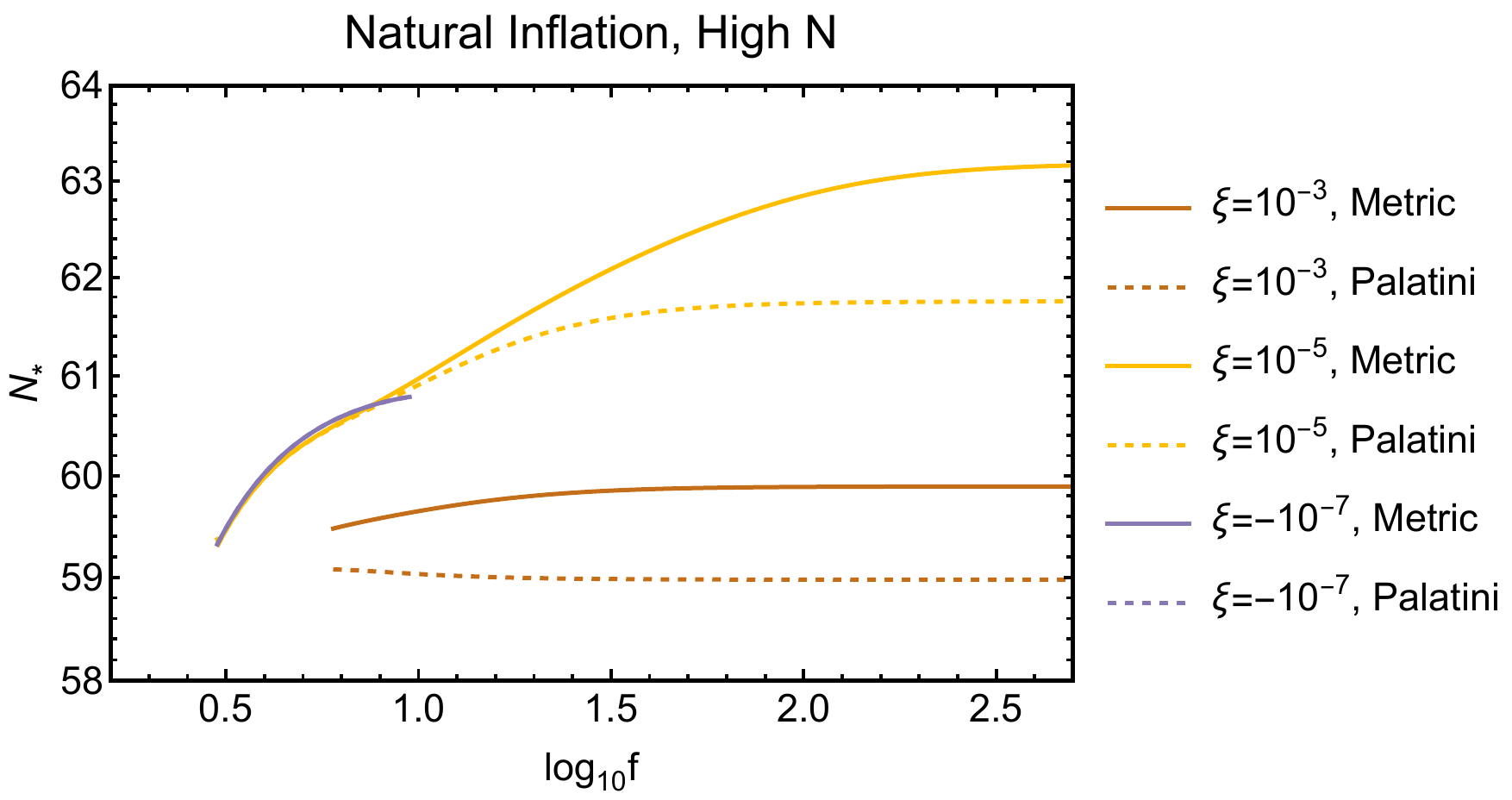}
\caption{\label{n42} For the Natural inflation potential, figures show the predictions of $n_s$, $r$, $N_*$ as a function of $f$ for chosen $\xi$ values, considering the high-$N$ case, both formulations and $n=4$. }
\end{figure}

\item Also, for $\xi>0$ values, $N_*$ remains constant while increasing $f$. For example, $N_* \sim 64(61)$ for $\xi=10^{-2}(10^{-3})$ and very large $f$ values. On the other hand, for $\xi=-10^{-2}$ and $\xi=-7\times 10^{-3}$ values, $N_*$ cannot stay as a constant. In addition, for $f\sim3$, $N_*$ is starting to decrease to $N_*\sim 58(56)$ in Metric formulation for $\xi=-10^{-2}(-7\times 10^{-3})$. Also, $N_*\sim 59$ for the Palatini approach for both $\xi=-10^{-2}$ and $\xi=-7\times 10^{-3}$ values.

\item For both $n=2$ and $n=4$ cases, we can emphasize that the predictions for the Metric and Palatini formulations are very close to each other for the weak coupling limit. For $\xi<0$ values, we observe that the inflationary predictions are in a better consistency with the recent data for the Palatini formulation than the Metric one. 

\item We also observe that the predictions are ruled out for the minimally-coupled ($\xi=0$) Natural Inflation potential with the latest constraints, in addition to this, for the conformal coupling ($\xi=-1/6$ in our metric convention), there are no solutions to satisfy the inflationary predictions ($n_s-r$) for this potential model.

\end{itemize}

\section{Conclusion}\label{conc}

In this work, we analyze the inflationary parameters of Natural Inflation potential with non-minimal coupling, $\xi\phi^2R$, between Ricci scalar and inflaton. We use a form of the non-minimal coupling function: $F(\phi)=1+\xi \phi^n$ for both Metric and Palatini approaches with the assumption of the standard thermal history after inflation. We show our results for $n=2$ and $n=4$ cases, we observe that the lowest ($n=2$) case gives a better agreement with the recent observations from BICEP/Keck for both two formulations. Also, for all selected $\xi$ values in this paper, the predictions are ruled out for $n=4$ case. In addition, we show that especially in the weak coupling limit, for both $n = 2$ and $n = 4$ cases, the predictions for both formulations are very close to each other.

We analyze the non-minimally coupled Natural Inflation potential in a very wide range of $f$ values, however, we find that for a very small range of $f$ can be inside the observational regions. By considering the low-$N$ case in the Palatini approach, the predictions can be inside the $95\%$ CL for $\xi=-10^{-2}$ values at $f\sim3.2$. Also, for high-$N$ case in the Metric and Palatini formulations, $n_s-r$ can enter to $95\%$ CL for the values of $\xi=-7\times 10^{-3}$ and $\xi=-10^{-2}$ at $f\sim 3.2$. We observe that when the inflationary predictions are inside the $95\%$ CL region, $N_* \sim 58$ for $\xi=-10^{-2}$ and $N_* \sim 56$ for $\xi=-7\times 10^{-3}$. Moreover, for $f\sim3$, in the high-$N$ case and Palatini approach, the inflationary predictions can be inside $68\%$ CL for $\xi=-10^{-2}$. In addition, we observe that as $T_{r}$ increases, inflationary predictions improve to be able to enter the observational regions, for example, we show that for the low-$N$ case and Metric formalism, the predictions are ruled out for $\xi=-7\times10^{-3}$, however, for the medium-$N$ case and Metric formalism, the predictions can stay inside the $95\%$ CL for the same $\xi$ value.

For $\xi>0$ values and $n=2$ case, when $\xi$ values increase, $n_s$ decreases and $r$ increases while increasing $f$, so $n_s-r$ values are naturally ruled out. So, in this work, we eliminate to show any predictions for the $\xi$ values larger than $10^{-2}$. This behavior is equivalent for both formulations. Also, for $\xi>0$ and $n=4$ case, the predictions do not change so much while increasing $f$, $n_s-r$ remains as a constant for $f\gg5$, $n_s\sim 0.95$ for all selected $\xi>0$ values. In addition, for $\xi=10^{-3}$, $r \sim 5\times 10^{-4}$ and for $\xi=10^{-5}$, $r \sim 0.04$. This behavior is also the same for both formulations. Moreover, for $\xi<0$ values, we observe that the inflationary predictions are better to be able to enter the observational regions for the Palatini formulation than the Metric one. In addition, we can emphasize that the agreement with the latest constraints is only possible with $\xi<0$ values for our model. As we already mentioned in the previous chapter, this might be considered a problem because $F(\phi)>0$ is required for any arbitrary value of $\phi$. However, we only take into account negative $\xi$ values that give $F>0$ during and after the inflationary era for our model.

Lastly, we find that the predictions are ruled out for the Natural Inflation potential with minimal coupling ($\xi=0$) with the latest constraints, as well as for the conformal coupling ($\xi=-1/6$ in our metric convention), there are no solutions to satisfy the inflationary predictions ($n_s-r$). Also, we can emphasize that the predictions of $\alpha$ are too small for the non-minimally coupled Natural Inflation potential to be observed in the future measurements.

\acknowledgments
I thank Antonio Racioppi, Durmu\c{s} Ali Demir, and Xinyi Zhang for a useful discussion.



\end{document}